\documentclass[]{article}
\usepackage{amsfonts}
\usepackage{amsthm,amssymb,amsmath}

\newtheorem*{theorem}{Theorem}

\newtheorem*{lemma}{Lemma}

\renewcommand{\theequation}{\thesection.\arabic{equation}}

\begin{document}

\author{B. Konopelchenko$^{a)}\,$\thanks{%
Supported in part by the grant COFIN 2000 ''Sintesi''.} , L.
Martinez
Alonso$%
^{b)}\,$\thanks{%
Supported in part by CICYT proyecto PB98-0821.} \\
\\
$^{a)}$Dipartimento di Fisica, Universita' di Lecce \\
and Sezione INFN, 73100 Lecce Italy; \\
$^{b)}$Departamento de Fisica Teorica II Universidad Complutense, \\
Madrid, Spain.}

\title{Dispersionless scalar integrable hierarchies, Whitham hierarchy and the
quasi-classical $\bar{\partial}$-dressing method.\thanks{This work
has been done in the framework of the grant INTAS-99-1782 }}

\maketitle

\begin{abstract}
The quasi-classical limit of the scalar nonlocal
$\bar{\partial}$-problem is derived and a quasi-classical version
of the $\bar{\partial}$-dressing method is presented.
Dispersionless KP, mKP and 2DTL hierarchies are discussed as
illustrative examples. It is shown that the universal Whitham
hierarchy it is nothing but the ring of symmetries for the
quasi-classical $\bar{\partial} $-problem. The reduction problem
is discussed and, in particular, the d2DTL equation of B type is
derived.
\end{abstract}

\section{Introduction}

\setcounter{equation}{0}

A considerable interest has been paid recently to dispersionless
or
quasi-classical limits of integrable equations and hierarchies (see \textit{%
e.g.} \cite{1}-\cite{13} and references therein). Study of
dispersionless hierarchies is of great importance since they arise
in the analysis of various problems in physics, mathematics and
applied mathematics from the theory of quantum fields and strings
\cite{14}-\cite{16} to the theory of conformal maps on the complex
plane \cite{17}-\cite{18}.

Different methods have been used to study dispersionless equations
and hierarchies \cite{1}-\cite{13}. In particular, several
1+1-dimensional equations and systems have been analyzed by the
quasi-classical version of the inverse scattering transform,
including the local Riemann-Hilbert problem approach
\cite{2},\cite{3},\cite{11}-\cite{13},\cite{19}. Similar study of
the 2+1-dimensional equations and hierarchies, like KP and 2DTL,
has been missing. Recently this problem has been addressed in
\cite{20} and the quasi-classical $\bar{\partial}$-dressing
approach to the dispersionless KP hierarchy has been proposed.

In this paper we consider a class of scalar dispersionless
integrable hierarchies governed by the scalar
$\bar{\partial}$-problem with the dKP, mdKP and d2DTL hierarchies
as particular cases. We derived the general form of the
quasi-classical $\bar{\partial}$-problem. It is given by the
system
\begin{eqnarray}
\frac{\partial S}{\partial \bar{\lambda}} &=&W\left( \lambda ,\bar{\lambda};%
\frac{\partial S}{\partial \lambda }\right),  \label{1.1} \\
\frac{\partial \varphi }{\partial \bar{\lambda}} &=&W^{\prime
}\left(
\lambda ,\bar{\lambda};\frac{\partial S}{\partial \lambda }\right) \frac{%
\partial \varphi }{\partial \lambda }+\widetilde{W}\left( \lambda ,%
\bar{\lambda};\frac{\partial S}{\partial \lambda }\right)
\frac{\partial ^{2}S}{\partial \lambda ^{2}}\,\varphi, \label{1.2}
\end{eqnarray}
for $\lambda \in G$, where $G$ is a domain in the complex plane
$\mathbb{C}$ and $W$ and $\widetilde{W}$ are some functions. The
type of hierarchy is specified by the undressed part
$S_{0}(\lambda ,T)$ of $S$ and the domain $G$. A quasiclassical
$\bar{\partial}$-dressing method based on the system
(\ref{1.1})-(\ref{1.2}) allows us to construct dispersionless
integrable hierarchies and provides us a method for finding their
solutions. The dKP, dmKP and d2DTL hierarchies are considered as
illustrative examples.

Symmetries of the quasi-classical $\bar{\partial}$-problem
(\ref{1.1})-(\ref {1.2}) are defined by linear Beltrami-type
equations and form an infinite-dimensional ring. It is shown that
this ring, parametrized by symmetry parameters, is nothing but the
universal Whitham hierarchy introduced in \cite{8}. In particular,
the dKP, mdKP and m2DTL hierarchies are special subrings of
symmetries for problems (\ref{1.1})-(\ref{1.2}).

We discuss also the reduction of the dispersionless hierarchies
and present the dispersionless 2DTL equation of B type.

Equations of the form (\ref{1.1})-(\ref{1.2}) are well-known in
the complex-analysis; in particular, in connection with
quasi-conformal mappings in the plane (see \textit{e.g.}
\cite{21}-\cite{23}). Thus, there is a close connection between
the theory of quasi-classical integrable hierarchies and the
theory of quasi-conformal mappings.

\section{Dispersionless hierarchies and
\\ %
universal Whitham hierarchy}

\setcounter{equation}{0}

We begin by reminding some relevant formulas for dispersionless
hierarchies and choose the Kadomtsev-Petviashili (KP) hierarchy to
illustrate their main features. The usual KP hierarchy is an
infinite set of the compatibility condition for the system
\begin{eqnarray}
L\psi &=&\lambda \psi  \label{2.1} \\
\frac{\partial \psi }{\partial t_{n}} &=&\left( L^{n}\right)
_{+}\psi \label{2.2}
\end{eqnarray}
where $L=\partial +u_{1}\partial ^{-1}+u_{2}\partial ^{-2}+...,\partial =%
\frac{\partial }{\partial t},\left( L^{n}\right) _{+}$ denotes the
pure differential part of the operator $L^{n}$, $\lambda $ is a
spectral parameter and $\psi $ is a common KP wave-function. The
KP equation itself is the equation for coefficient $u_{1}$ as a
function of the first three
times $t_{1},t_{2},t_{3}$. For the modified KP (mKP) hierarchy the operator $%
L$ is of the form $L=\partial +u_{0}+u_{1}\partial
^{-1}+u_{2}\partial ^{-2}+...$ while for the two-dimensional Toda
lattice (2DTL) hierarchy one needs two operators $L_{1}$ and
$L_{2}$ \cite{10}.

The dispersionless KP (dKP) hierarchy is a formal limit
$\varepsilon\rightarrow 0$ of the KP hierarchy for which
\cite{1}-\cite{10}
\begin{equation}
u_{k} \left( \frac{T_{n}}{\varepsilon }\right)
\mathop{\longrightarrow}_{\varepsilon\rightarrow 0}
u_{k_{0}}(T)+O(\varepsilon ) , \label{2.3}
\end{equation}
and
\begin{equation}
\psi \left( \frac{T_{n}}{\varepsilon }\right)
\mathop{\longrightarrow}_{\varepsilon\rightarrow 0}
e^{\frac{1}{\varepsilon }S(\lambda,T)+O(\varepsilon )},
\label{2.4}
\end{equation}
where $T_{n}$ are slow times.

Under such a limit, equation (\ref{2.1}) gives rise to the Laurent
series
$\cal L$%
$=p+\sum_{n=1}^{\infty }u_{n}(T)\,p^{-n}$, where $p=\frac{\partial S}{%
\partial T_{1}}$ while equations (\ref{2.2}) become
\begin{equation}
\frac{\partial p}{\partial T_{n}}=\frac{\partial
B_{n}(p)}{\partial T_{1}} \label{2.5}
\end{equation}
where $B_{n}(p)=[$%
$\cal L$%
$^{n}(p)]_{+}$ and $[$%
$\cal L$%
$^{n}]_{+}$ denotes here a polynomial part of
$\cal L$%
$^{n}$. The compatibility conditions for (\ref{2.5}) are given by
the infinite set of equation
\begin{equation}
\frac{\partial B_{n}}{\partial T_{m}}-\frac{\partial B_{n}}{\partial T_{m}}%
+\{B_{n},B_{m}\}=0  \label{2.6}
\end{equation}
where the Poisson bracket $\{,\}$ is defined as
\begin{equation}
\{f,g\}=\frac{\partial f}{\partial p}\,\,\frac{\partial g}{\partial T_{1}}-\,%
\frac{\partial f}{\partial T_{1}}\,\,\frac{\partial g}{\partial
p}\qquad . \label{2.7}
\end{equation}
Equation (\ref{2.5}) or (\ref{2.6}) represent the dKP hierarchy.
Similarly, the dmKP hierarchy is given by equations of the form
(\ref{2.5})-(\ref{2.7}) with
$\cal L$%
$=p+\sum_{n=1}^{\infty }u_{n}(T)\,p^{-n}$ \cite{25}. The d2DTL
hierarchy can be written by a set of equations similar to
(\ref{2.5})-(\ref{2.7}) for two Laurent series
$\cal L$%
$_{1}$ and
$\cal L$%
$_{2}$ \cite{10} with the substitution $p\rightarrow e^{p}$.

A more general dispersionless-like hierarchy has been introduced
in \cite{8}. This universal Whitham hierarchy is given by the
infinite set of equations
\begin{equation}
\frac{\partial \Omega _{A}}{\partial T_{B}}-\frac{\partial \Omega _{B}}{%
\partial T_{A}}+\{\Omega _{A},\Omega _{B}\}=0\quad \quad ,\qquad
A,B=1,2,3,...  \label{2.8}
\end{equation}
where $\Omega _{A}(p,T)$ are arbitrary holomorphic functions of
$p$. As it has been shown in \cite{8}, the dKP, d2DTL  and Benney
hierarchies are particular cases.

\section{Quasi-classical $\bar{\partial}$-problem}

\setcounter{equation}{0}

The $\vec{\partial}$-dressing method is a powerful tool to study
usual integrable equations and hierarchies \cite{25}-\cite{27}. In
this paper we shall formulate its quasi-classical version. We
shall demonstrate that it provides an effective method to
construct and study dispersionless hierarchies.

We begin with the derivation of the quasi-classical limit of the basic $\bar{%
\partial}$-problem.

The usual scalar integrable hierarchies are associated with the
following scalar linear nonlocal problem (see \cite{25}-\cite{27})
\begin{equation}
\frac{\partial \,\chi (\lambda ,\overline{\lambda };t )}{\partial \,%
\overline{\lambda }}=\int \int_{\mathbb{C}}d\mu \wedge
d\overline{\mu }\,\,\chi
(\mu ,\overline{\mu };t)\,\,g(\mu ,t)\,R_{0}(\mu ,\overline{\mu };\lambda ,%
\overline{\lambda })\,g^{-1}(\lambda ,t)  \label{3.1}
\end{equation}
where $\lambda $ is a complex variable (''spectral parameter''), $\overline{%
\lambda }$ denotes complex conjugation of $\lambda ,\chi (\lambda ,\overline{%
\lambda };\mu )$ is a complex-valued function on the complex plane
$\mathbb{C}$
($\lambda ,\overline{\lambda }\in \mathbb{C}$), the kernel $R_{0}(\mu ,%
\overline{\mu };\lambda ,\overline{\lambda })$ is the
$\bar{\partial}$-data. Usually, it is assumed that the function
$\chi $ has a canonical
normalization (\textit{i.e.}
\[
\chi \rightarrow 1+\frac{\chi _{1}}{\lambda }+%
\frac{\chi _{2}}{\lambda ^{2}}+...,\quad \lambda \rightarrow
\infty,
\]
and that the problem (\ref{3.1}) is uniquely solvable. Concrete
integrable hierarchies are specialized by the form of the function
$g(\lambda ,t)=\exp
(S_{0}(\lambda ,t))$ and by the domain $G$ of the support for the $\bar{%
\partial}$-data $R_{0}$ ($R_{0}(\mu ,\overline{\mu };\lambda ,\overline{%
\lambda })\,=0$ for $\mu ,\lambda \in $
$\mathbb{C}/G$). For the KP hierarchy $%
S_{0}=\sum_{k=1}^{\infty }\lambda ^{k}t_{k}$ and $G$ ia a disk
with center at the origin,  while for the 2DTL hierarchy
$S_{0}(\lambda ;x,y,n)=n\ln \lambda+\sum_{k=1}^{\infty }\lambda
^{k}x_{k}+\sum_{k=1}^{\infty }\lambda ^{-k}y_{k}$ where $x_{k}$
and $y_{k}$ are continuous variables and $n$ is an integer
discrete variable. The domain
$G$ in this case is an annulus  $a\leq |\lambda |\leq b$. Given $%
g(\lambda )$ the $\bar{\partial}$-dressing method provides us with
the corresponding hierarchy of nonlinear equations and their
linear problems \cite{25}-\cite{27}. Solutions of nonlinear
equations are given by the function $\chi $ evaluated at certain
points $\lambda _{0}$. For instance, for the KP hierarchy
$u_{1}=-2\frac{\partial \chi _{1}(t)}{\partial t_{1}}$ .

In order to derive the quasi-classical limit of the
$\bar{\partial}$-problem
(\ref{3.1}) we first introduce slow variables $T$ ($t_{i}=\frac{T_{i}}{%
\varepsilon }\,$for KP and mKP, $x_{i}=\frac{X_{i}}{\varepsilon }$, $y_{i}=%
\frac{Y_{i}}{\varepsilon }$, $n=\frac{T}{\varepsilon }$ for 2DTL ) for small $%
\varepsilon $ and proceed to the limit $\varepsilon \rightarrow
0$. In
this limit $g\left( \frac{T}{\varepsilon }\right) =\exp \left[ \frac{%
S_{0}(\lambda ,T)}{\varepsilon }\right] $ . Motivated by the
formula of the type (\ref{2.4}) and by the structure of equation
(\ref{3.1}) we will look for solutions $\chi $ of the form
\begin{equation}
\chi \left( \lambda ,\overline{\lambda };\frac{T}{\varepsilon }\right) =%
\widehat{\chi }(\lambda ,\overline{\lambda };T;\varepsilon )\,\,e^{\frac{%
\widetilde{S}(\lambda ,\overline{\lambda };T)}{\varepsilon }}
\label{3.2}
\end{equation}
where $\widetilde{S}(\lambda ,\overline{\lambda };T)$ is a certain
function and
\begin{equation}
\widehat{\chi }(\lambda ,\overline{\lambda };T;\varepsilon
)=\sum_{n=0}^{\infty }\widehat{\chi }_{n}(\lambda ,\overline{\lambda }%
;T)\,\varepsilon ^{n}\qquad .  \label{3.3}
\end{equation}
It is clear that only for special $\bar{\partial}$-data $R_{0}$ equation (%
\ref{3.1}) will have a well-defined limit as $\varepsilon
\rightarrow 0$. Thus is not difficult to see that the $\bar{\partial}$%
-data of the form
\begin{equation}
R_{0}(\mu ,\overline{\mu };\lambda ,\overline{\lambda
};\varepsilon )\,=\sum_{k=0}^{\infty }\Gamma _{k}(\mu
,\overline{\mu })\,\varepsilon
^{k-1}\,\delta ^{(k)}(\mu -\lambda -\varepsilon \,\alpha _{k}(\lambda ,%
\overline{\lambda }))  \label{3.4}
\end{equation}
do a job. Here $\Gamma _{k}(\mu ,\overline{\mu }),\;\alpha _{k}(\lambda ,%
\overline{\lambda })$ are arbitrary functions ($\Gamma _{k}=0$ at
$\lambda \in \mathbb{C}/G$) and $\delta ^{(k)}$ is the
$k$-derivative Dirac delta-function. Indeed, substituting
(\ref{3.4}) into (\ref{3.1}), one gets
\[
\frac{\partial \widehat{\chi }(\lambda ,\overline{\lambda
};T,\epsilon)}{\partial
\overline{\lambda }}+\frac{1}{\varepsilon }\frac{\partial S(\lambda,\overline{\lambda };T)}{%
\partial \overline{\lambda }}\;\widehat{\chi }(\lambda ,\overline{\lambda }%
,T,\epsilon)=\qquad \qquad \qquad \qquad \qquad \qquad \qquad
\qquad
\]
\begin{eqnarray}
&=&\int \int_{\mathbb{C}}d\mu \wedge d\overline{\mu
}\;\widehat{\chi }(\mu
,%
\overline{\mu },\epsilon)\;e^{\frac{S(\mu ,T)-S(\lambda ,T)}{\varepsilon }%
}\,\sum_{k=0}^{\infty }\Gamma _{k}(\mu ,\overline{\mu }%
)\,\varepsilon ^{k-1}\delta ^{(k)}(\mu -\lambda -\varepsilon
\alpha
_{k}(\lambda ,\overline{\lambda })\,)  \nonumber \\
&&  \label{3.5}
\end{eqnarray}
where
\[
S(\lambda ,\overline{\lambda };T):=\widetilde{S}(\lambda ,\overline{%
\lambda };T)+S_{0}(\lambda ;T).
\]
Evaluating in (\ref{3.5}) the terms of the order
$\frac{1}{\varepsilon }$, one obtains
\begin{equation}
\frac{\,\partial S(\lambda ,\overline{\lambda };T)}{\partial
\overline{\lambda }}=W\left( \lambda ,\overline{\lambda };\frac{\partial S}{%
\partial \lambda }\right)  \label{3.6}
\end{equation}
where
\begin{equation}
W\left( \lambda ,\overline{\lambda };\frac{\partial S}{\partial \lambda }%
\right) =\sum_{k=0}^{\infty }(-1)^{k}\,\Gamma _{k}(\lambda ,\overline{%
\lambda })\,\left( \frac{\partial S}{\partial \lambda }\right)
^{k}\,\,e^{\alpha _{k}(\lambda ,\overline{\lambda })\,\frac{\partial S}{%
\partial \lambda }}\qquad .  \label{3.7}
\end{equation}
Furthermore, the terms of zero order in $\varepsilon $ in
(\ref{3.5}) give (the contribution proportional to $\widehat{\chi
}_{1}$ disappears due to (\ref {3.6})):
\begin{equation}
\frac{\partial \varphi }{\partial \overline{\lambda }}=W^{\prime
}\left(
\lambda ,\overline{\lambda };\frac{\partial S}{\partial \lambda }\right) \,%
\frac{\partial \varphi }{\partial \lambda }+\widetilde{W}\left( \lambda ,%
\overline{\lambda };\frac{\partial S}{\partial \lambda }\right) \,\frac{%
\partial ^{2}S}{\partial \lambda ^{2}}\varphi  \label{3.8}
\end{equation}
where
\[
\varphi: =\widehat{\chi }_{0},\quad W^{\prime }(\lambda ,\overline{\lambda }%
;\xi )\,:=\frac{\partial W(\lambda ,\overline{\lambda };\xi
)\,}{\partial \xi },
\]
and
\begin{eqnarray}
\widetilde{W} &:=&(-1)\,\Gamma _{1}\,e^{\alpha _{1}\frac{\partial
S}{\partial
\lambda }}\frac{1}{2}\alpha _{1}^{2}\frac{\partial S}{\partial \lambda }%
+(-1)^{2}\,\Gamma _{2}\,e^{\alpha _{2}\frac{\partial S}{\partial \lambda }%
}\left[ 1+\frac{1}{2}\alpha _{2}^{2}\left( \frac{\partial
S}{\partial
\lambda }\right) ^{2}\right] +  \nonumber \\
&&+(-1)^{3}\,\Gamma _{3}\,e^{\alpha _{3}\frac{\partial S}{\partial \lambda }%
}\left[ 3\frac{\partial S}{\partial \lambda }+\frac{1}{2}\alpha
_{3}^{2}\left( \frac{\partial S}{\partial \lambda }\right)
^{3}\right] +... \label{3.9}
\end{eqnarray}
Since  $\frac{\,\partial S_{0}}{\partial \overline{\lambda }}=0$
at $\lambda \in G$,  then the equations (\ref{3.6})-(\ref{3.9})
for $\lambda \in G$ can be rewritten as
\begin{eqnarray}
\frac{\partial S}{\partial \overline{\lambda }} &=&W\left( \lambda ,%
\overline{\lambda };\frac{\partial S}{\partial \lambda }\right),
\label{3.10}
\\
\frac{\partial \varphi }{\partial \overline{\lambda }} &=&W^{\prime }\frac{%
\partial \varphi }{\partial \lambda }+\widetilde{W}\,\frac{\partial ^{2}S}{%
\partial \lambda ^{2}}\,\varphi .
\label{3.11}
\end{eqnarray}
Equations (\ref{3.10}) and (\ref{3.11}) are the quasi-classical
limit of the nonlocal $\bar{\partial}$-problem (\ref{3.1}). The
derivation given above suggests that the quasiclassical limit of
the $\bar{\partial}$-problem (\ref {3.1}) is given by equations
(\ref{3.10}), (\ref{3.11}) also for a more general than
(\ref{3.4}) $\bar{\partial}$-data $R_{0}$.

The function $S$ is widely used in the analysis of the
dispersionless limits
of the integrable hierarchies \cite{4}-\cite{10}. Within the $\bar{\partial}$%
-approach it is a nonholomorphic function of the ''spectral'' variable $%
\lambda $ and obeys the nonlinear $\bar{\partial}$-equation
(\ref{3.10}) (for $\lambda \in G$). The function $\varphi
=\widehat{\chi }_{0}$ obeys the local $\bar{\partial}$-problem
(\ref{3.11}) of the Beltrami type. Note that the ratio $\phi $ of
two solutions $\varphi _{1}$ and $\varphi _{2}$ of equation
(\ref{3.11}) satisfies the Beltrami equation
\begin{equation}
\frac{\partial \phi }{\partial \overline{\lambda }}=W^{\prime
}\left(
\lambda ,\overline{\lambda };\frac{\partial S}{\partial \lambda }\right) \,\,%
\frac{\partial \phi }{\partial \lambda }.  \label{3.12}
\end{equation}
For the Orlov's function $M=\frac{\partial S}{\partial \lambda }$ equations (%
\ref{3.10}) and (\ref{3.11}) take the form of quasi-linear
equations
\begin{eqnarray}
\frac{\partial M}{\partial \overline{\lambda }} &=&\frac{\partial
}{\partial \lambda }W\left( \lambda ,\overline{\lambda };M\right)
,  \label{3.13}
\\
\frac{\partial \varphi }{\partial \overline{\lambda }}
&=&W^{\prime }\left( \lambda ,\overline{\lambda };M\right)
\frac{\partial \varphi }{\partial
\lambda }+\widetilde{W}\,\left( \lambda ,\overline{\lambda };M\right) \frac{%
\partial M}{\partial \lambda }\,\varphi .  \label{3.14}
\end{eqnarray}
In the particular case of the $\bar{\partial}$-data $R_{0}$ given
by (\ref
{3.4}) with all $\alpha _{k}\equiv 0$ one has $\widetilde{W}\,=\frac{1}{2}%
W^{\prime \prime }\left( \lambda ,\overline{\lambda };\frac{\partial S}{%
\partial \lambda }\right) $.

Quasi-classical $\bar{\partial}$-problems (\ref{3.10}),
(\ref{3.11}) are
basic equations for our approach. The equations of the type (\ref{3.10}), (%
\ref{3.11}) are well known and widely studied in the theory of
nonlinear elliptic systems with two independent variables and in
complex analysis (see \textit{e.g.
}\cite{28},\cite{21}-\cite{23}). One theorem from the theory of
such equations will be crucial for our further constructions. This
theorem (see theorem 3.32 from \cite{28}) states that , under certain mild condition on $A$%
(see the appendix), the only solution of the
Beltrami equation $\frac{\partial Z}{\partial \overline{\lambda }}=A\frac{%
\partial Z}{\partial \lambda }$ in $\mathbb{C\,}$ which vanish as
$\lambda \rightarrow \infty$ is $Z\equiv 0$.

\section{Quasi-classical $\bar{\partial}$-dressing method}

\setcounter{equation}{0}

The principal goal of the $\bar{\partial}$-dressing method based
on
equations (\ref{3.10}), (\ref{3.11}) is the same as of the original $\bar{%
\partial}$-dressing method \cite{25}-\cite{27}. It is to extract the
nonlinear differential equations from the quasi-classical $\bar{\partial}$%
-problems.

Now the time dependence of the functions $S$ and $\varphi $ is
encoded in
the undressed functions $S_{0}(\lambda ,T)$. Since $\widetilde{S}=1+\frac{%
S_{1}}{\lambda }+\frac{S_{2}}{\lambda ^{2}}+...$ at $\lambda
\rightarrow
\infty $ then the behavior of $\frac{\partial S}{\partial T_{A}}$ for large $%
\lambda $ is completely defined by
\begin{equation}
\frac{\partial S}{\partial T_{A}}=\frac{\partial S_{0}}{\partial T_{A}}+%
\frac{1}{\lambda }\,\frac{\partial S_{1}}{\partial T_{A}}+...
\label{4.1}
\end{equation}
where $T_{A}$ is a slow time.

A basic property of the nonlinear equation (\ref{3.10}) is that it
implies the linear Beltrami equation for the infinitesimal
variations $\delta S$ (symmetries):
\begin{equation}
\frac{\partial }{\partial \overline{\lambda }}(\delta S)=W^{\prime
}\left(
\lambda ,\overline{\lambda };\frac{\partial S}{\partial \lambda }\right) \,\,%
\frac{\partial }{\partial \lambda }(\delta S)\qquad .  \label{4.2}
\end{equation}
In particular, for any time $T_{A}$
\begin{equation}
\frac{\partial }{\partial \overline{\lambda }}\left( \frac{\partial S}{%
\partial T_{A}}\right) =W^{\prime }\left( \lambda ,\overline{\lambda };\frac{%
\partial S}{\partial \lambda }\right) \,\,\frac{\partial }{\partial \lambda }%
\left( \frac{\partial S}{\partial T_{A}}\right) \qquad .
\label{4.3}
\end{equation}
Any power of solution of the Beltrami equation is a solution too
as well as
any differentiable function of two solutions. So together with $\frac{%
\partial S}{\partial T_{A_{1}}},...,\frac{\partial S}{\partial T_{A_{n}}}$
any differentiable function $f\left( \frac{\partial S}{\partial T_{A_{1}}}%
,...,\frac{\partial S}{\partial T_{A_{n}}}\right) $ with arbitrary
$n$ is a solution of equation (\ref{4.3}). Thus the symmetries of
the problem (\ref {3.10}) form a ring.

Due to (\ref{4.1}), the functions $f\left( \frac{\partial
S}{\partial T_{A_{1}}},...,\frac{\partial S}{\partial
T_{A_{n}}}\right) $ have
singularities in certain points. The functions $f_{0}\left( \frac{\partial S%
}{\partial T_{A_{1}}},...,\frac{\partial S}{\partial
T_{A_{n}}}\right) $ which are bounded in $\mathbb{C}$ and vanish
as $\lambda \rightarrow \infty $ are very special. According to
the Vekua's theorem mentioned in the end of the previous section
they vanish identically. So we have the nonlinear equations
\begin{equation}
f_{0}\left( \frac{\partial S}{\partial T_{A_{1}}},...,\frac{\partial S}{%
\partial T_{A_{n}}}\right) =0\qquad .  \label{4.4}
\end{equation}
Note that in contrast to the usual hierarchies we get nonlinear
equations for the ''wave function'' $S$ (the classical action).

Now we turn to the $\bar{\partial}$-problem (\ref{3.11}). It is
linear one. So the construction is similar to that of usual case.
Namely, suppose that one has found a solution $\varphi _{0}$ to
equation (\ref{3.11}) of the form $\varphi _{0}=L\varphi $ where
$L$ is a certain linear operator and $\varphi
_{0}$ is bounded and vanishes as $\lambda \rightarrow \infty $. Since $%
\varphi \rightarrow 1+\frac{\varphi _{1}}{\lambda}+...\,$as
$\lambda \rightarrow \infty $ the ratio $\frac{\varphi
_{0}}{\varphi }$ vanishes as $\lambda \rightarrow \infty $ and
obeys the Beltrami equation
\begin{equation}
\frac{\partial }{\partial \overline{\lambda }}\left( \frac{\varphi _{0}}{%
\varphi }\right) =W^{\prime }\,\frac{\partial }{\partial \lambda
}\left( \frac{\varphi _{0}}{\varphi }\right) \qquad .  \label{4.5}
\end{equation}
Then according to the Vekua's theorem $\frac{\varphi _{0}}{\varphi
}$ vanishes identically and, consequently,
\begin{equation}
L\varphi =0  \label{4.6}
\end{equation}
that is the desired linear problem for the wavefunction $\varphi
$. Note that one can get the same results assuming that the
problem (\ref{3.11}) with canonically normalized $\varphi $ is
uniquely solvable.

Equations (\ref{4.4}) and (\ref{4.6}) are the basic equations
associated
with the quasi-classical $\bar{\partial}$-problems (\ref{3.10}), (\ref{3.11}%
). They are compatible by construction. Equation (\ref{4.4}),
(\ref{4.6}) provide us also with equations for functions
$u_{k}(T)$ which depend only on
the times $T$. Usually one has infinite families of equations of the type (%
\ref{4.4}), (\ref{4.6}). So the quasi-classical $\bar{\partial}$-problems (%
\ref{3.10}), (\ref{3.11}) give rise to an infinite hierarchy of
integrable quasi-classical (or dispersionless) equations.

\section{dKP and dmKP hierarchies}

\setcounter{equation}{0}

Let us consider concrete examples to illustrate the general
scheme. We start with the dKP hierarchy. In this case
$S_{0}(\lambda ,T)=\sum_{n=1}^{\infty
}\lambda ^{n}T_{n}$ and $\frac{\partial S}{\partial T_{n}}=\lambda ^{n}+%
\frac{1}{\lambda }\frac{\partial S_{1}}{\partial T_{n}}+\frac{1}{\lambda ^{2}%
}\frac{\partial S_{2}}{\partial T_{n}}+...$ ($n=1,2,....)$ . Since $\frac{%
\partial S}{\partial T_{n}}$ have power singularities at infinity the
desired function $f_{0}$ will be, clearly, polynomials. Taking,
for
instance, the derivatives $\frac{\partial S}{\partial T_{2}}$ and $\frac{%
\partial S}{\partial T}$ we readily see that the difference $\frac{\partial S%
}{\partial T_{2}}-\left( \frac{\partial S}{\partial T_{1}}\right)
^{2}$ behaves as $-2\frac{\partial S_{1}}{\partial
T_{1}}+O(\frac{1}{\lambda })$ as $\lambda \rightarrow \infty $.
Thus, the desired function $f_{0}$ is  $\frac{\partial S}{\partial T_{2}}-\left( \frac{\partial S}{\partial T_{1}}%
\right) ^{2}+2\frac{\partial S_1}{\partial T_{1}}$. So we get the
equation
\begin{equation}
\frac{\partial S}{\partial T_{2}}-\left( \frac{\partial S}{\partial T_{1}}%
\right) ^{2}-u=0  \label{5.1}
\end{equation}
where $u=-2\frac{\partial S_{1}}{\partial T_{1}}$ .

Analogously, taking the derivatives $\frac{\partial S}{\partial T_{3}}$ and $%
\frac{\partial S}{\partial T_{1}}$ one easily concludes that the
combination
$\frac{\partial S}{\partial T_{3}}-\left( \frac{\partial S}{\partial T_{1}}%
\right) ^{3}-V_{1}\frac{\partial S}{\partial T_{1}}-V_{0}$ vanishes at $%
\lambda \rightarrow 0$ if $V_{1}=-3\frac{\partial S}{\partial T_{1}}$ and $%
V_{0}=-3\frac{\partial S_{2}}{\partial
T_{1}}=-\frac{3}{2}\frac{\partial S_{1}}{\partial T_{2}}$ . So one
gets the other function $f_{0}$ and the equation
\begin{equation}
\frac{\partial S}{\partial T_{3}}-\left( \frac{\partial S}{\partial T_{1}}%
\right) ^{3}-V_{1}\frac{\partial S}{\partial T_{1}}-V_{0}=0
\label{5.2}
\end{equation}
where $\frac{\partial V_{0}}{\partial T_{1}}=\frac{3}{4}\frac{\partial u}{%
\partial T_{2}}$ .

In a similar manner one constructs an infinite family of equations
\begin{equation}
\frac{\partial S}{\partial T_{n}}-\left( \frac{\partial S}{\partial T_{1}}%
\right) ^{n}-\sum_{k=0}^{n-2}V_{nk}(T)\,\,\left( \frac{\partial
S}{\partial T_{1}}\right) ^{k}=0\qquad ,\quad n=1,2,3,...
\label{5.3}
\end{equation}
with appropriate coefficients $V_{nk}(T)$.

Equations (\ref{5.3}) are nothing but the equations (\ref{2.5}) of
the dKP hierarchy. Evaluating the left-hand-sides of (\ref{5.3})
at $\lambda \rightarrow \infty $, taking the term of the order
$\frac{1}{\lambda }$ and using other equations (\ref{5.3}), one
gets the dKP hierarchy for the function $u$ and, in particular,
the dKP equation
\begin{equation}
u_{T_{1}T_{3}}=\frac{3}{2}(u\,u_{T_{1}})_{T_{1}}+\frac{3}{4}%
u_{T_{2}T_{2}}\qquad .  \label{5.4}
\end{equation}
In a usual manner equation (\ref{5.4}) arises as the compatibility
conditions for equations (\ref{5.1}) and (\ref{5.2}). Equations
(\ref{5.3}) implies the hierarchy of nonlinear equations for the
function $S$ only. Indeed, eliminating all coefficients
$u_{nk}(T)$ from (\ref{5.3}) one gets the family of equations
\[
\frac{\partial S}{\partial T_{n}}=F_{n}\left( \frac{\partial
S}{\partial T_{1}},\frac{\partial S}{\partial T_{2}}\right) \qquad
\quad ,\qquad n=3,4,5,...
\]
The lowest of these equations is of the form
\[
\frac{\partial ^{2}S}{\partial T_{1}\partial T_{3}}=\frac{3}{4}\frac{%
\partial ^{2}S}{\partial T_{2}^{2}}+\frac{3}{2}\left[ \frac{\partial S}{%
\partial T_{2}}-\left( \frac{\partial S}{\partial T_{1}}\right) ^{2}\right]
\,\,\frac{\partial ^{2}S}{\partial T_{1}^{2}}\qquad .
\]
Now we proceed to the $\bar{\partial}$-problem (\ref{3.11}). For
simplicity we restrict ourselves to the case
$\widetilde{W}=\frac{1}{2}W^{\prime \prime }$. It is not difficult
to show, differentiating (\ref{3.11}) and using
equations (\ref{5.1})-(\ref{5.2}), that the function $Z=L\varphi =\frac{%
\partial \varphi }{\partial T_{2}}-2\frac{\partial S}{\partial T_{1}}\frac{%
\partial \varphi }{\partial T_{1}}-\frac{\partial ^{2}S}{\partial T_{1}^{2}}%
\varphi +\widetilde{u}\varphi $ where
$\widetilde{u}=-2\frac{\partial \varphi _{1}}{\partial T_{1}}$
obeys the equation
\begin{equation}
\frac{\partial Z}{\partial \overline{\lambda }}=W^{\prime }\,\frac{\partial Z%
}{\partial \lambda }+\frac{1}{2}W^{\prime \prime }\frac{\partial ^{2}S}{%
\partial \lambda ^{2}}Z  \label{5.5}
\end{equation}
and vanish at $\lambda \rightarrow 0$ . Consequently the ratio $\frac{Z}{%
\varphi }$ obeys the Beltrami equation $\frac{\partial }{\partial \overline{%
\lambda }}\left( \frac{Z}{\varphi }\right) =W^{\prime }\,\frac{\partial }{%
\partial \lambda }\left( \frac{Z}{\varphi }\right) $ and $\frac{Z}{\varphi }%
=O(\frac{1}{\lambda })$ as $\lambda \rightarrow \infty $.
According to the Vekua's theorem this ratio vanishes identically
and, consequently, we get the linear problem $Z=0$, \textit{i.e.}
\begin{equation}
\frac{\partial \varphi }{\partial T_{2}}-2\frac{\partial S}{\partial T_{1}}%
\frac{\partial \varphi }{\partial T_{1}}-\frac{\partial
^{2}S}{\partial T_{1}^{2}}\varphi +\widetilde{u}\varphi =0\qquad .
\label{5.6}
\end{equation}
In a similar manner, one gets the equation
\begin{eqnarray}
\frac{\partial \varphi }{\partial T_{3}}-6\left[ 2\left( \frac{\partial S}{%
\partial T_{1}}\right) ^{2}+\widetilde{u} \right] \frac{\partial \varphi }{%
\partial T_{1}}-3\left[ 4\frac{\partial S}{\partial T_{1}}\frac{\partial
^{2}S}{\partial T_{1}^{2}}+\frac{\partial \widetilde{u} }{\partial T_{1}}-2%
\widetilde{u}\frac{\partial S}{\partial
T_{1}}+\widetilde{w}\right] \varphi
&=&0  \nonumber \\
&&  \label{5.7}
\end{eqnarray}
and higher-time equations
\begin{equation}
\frac{\partial \varphi }{\partial T_{n}}-A_{n}\frac{\partial \varphi }{%
\partial T_{1}}-B_{n}\varphi =0\qquad ,\qquad n=1,2,3...  \label{5.8}
\end{equation}
All linear problems (\ref{5.6})-(\ref{5.8}) are compatible by
construction.

In the particular case $\widetilde{u}=\widetilde{w}=0$ equation
(\ref{5.6}) is known as the transport equation within the
quasiclassical approximation in quantum mechanics (see
\textit{e.g.} [29]). Note that in the case
$\widetilde{u}=\widetilde{w}=0$  equations (\ref{5.6}), (\ref{5.7}) (and also (%
\ref{5.8}))  take the form of conservation laws
\begin{equation}
\begin{array}{l}
\frac{\partial \phi }{\partial T_{2}}-\frac{\partial }{\partial
T_{1}}\left(
\frac{\partial S}{\partial T_{1}}\phi \right) =0\qquad , \\
\frac{\partial \phi }{\partial T_{3}}-\frac{\partial }{\partial T_{1}}%
\Big( 12\Big( \frac{\partial S}{\partial T_{1}}\Big) ^{2} \phi
\Big) =0\qquad .
\end{array}
\label{5.9}
\end{equation}

Considering the adjoint dKP hierarchy for which $\psi ^{*}(T)=\widetilde{%
\chi }^{*}(T,\lambda ;\varepsilon )e^{-\frac{S}{\varepsilon }}$,
one gets the same equation (\ref{4.1}) for $S$ and equations for
$\varphi ^{*}$ which
are adjoint to (\ref{5.6})-(\ref{5.7}). It is interesting that the quantity $%
\varphi (\lambda ,T)\,\varphi ^{*}(\lambda ,T)$ obeys exactly
equations (\ref {5.9}).

Our second example is given by the dmKP hierarchy. In this case $%
S_{0}=\sum_{k=1}^{\infty }\lambda ^{-k}T_{k}$ and $\frac{\,\partial S}{%
\partial T_{k}}=\frac{1}{\lambda ^{k}}+\frac{\partial \widetilde{S}(\lambda
,T)}{\partial T_{k}}$ where $\widetilde{S}(\lambda ,T)$ is
holomorphic around $\lambda =0$. So to construct required
functions $f_{0}$ one has to
cancel singularities around $\lambda =0$. Taking again the derivatives $%
\frac{\partial S}{\partial T_{1}}$ and $\frac{\partial S}{\partial
T_{2}}$
one readily see that the combination $\frac{\partial S}{\partial T_{2}}%
-\left( \frac{\partial S}{\partial T_{1}}\right) ^{2}$ has only
simple pole at $\lambda =0$. To cancel it, we subtract
$V(T)\,\frac{\partial S}{\partial T_{1}}$ where
$V(T)\,=-2\frac{\partial
\widetilde{S}(\lambda =0)}{\partial T_{2}}$ . Then at $\lambda =0$ one has $%
\frac{\partial S}{\partial T_{2}}-\left( \frac{\partial S}{\partial T_{1}}%
\right) ^{2}-V(T)\,\frac{\partial S}{\partial T_{1}}=O\left( \frac{1}{%
\lambda }\right) $. So due to the Vekua's theorem we conclude
\begin{equation}
\frac{\partial S}{\partial T_{2}}-\left( \frac{\partial S}{\partial T_{1}}%
\right) ^{2}-V(T)\,\frac{\partial S}{\partial T_{1}}=0
\label{5.10}
\end{equation}
where $V(T)=-2\frac{\partial \widetilde{S}(\lambda =0,T)}{\partial
T_{2}}$ .
Taking the derivatives $\frac{\partial S}{\partial T_{1}}$ and $\frac{%
\partial S}{\partial T_{3}}$, one finds the equation
\begin{equation}
\frac{\partial S}{\partial T_{3}}-\left( \frac{\partial S}{\partial T_{1}}%
\right) ^{3}-\frac{3}{2}V\,\,\,\left( \frac{\partial S}{\partial T_{1}}%
\right) ^{2}-\left( \frac{3}{4}V^{2}-3W\right) \frac{\partial
S}{\partial T_{1}}=0  \label{5.11}
\end{equation}
where $\widetilde{S}(\lambda )=\widetilde{S}(0)+\lambda W(T)+...$ as $%
\lambda \rightarrow 0$. Analogously, one obtain the infinite
hierarchy of equations
\begin{equation}
\frac{\partial S}{\partial T_{n}}-\left( \frac{\partial S}{\partial T_{1}}%
\right) ^{n}-\sum_{k=1}^{n-1}V_{nk}(T)\,\,\,\left( \frac{\partial S}{%
\partial T_{1}}\right) ^{k}=0\qquad ,\qquad n=1,2,3...  \label{5.12}
\end{equation}
Equations (\ref{5.12}) give us the dmKP hierarchy (see
\textit{e.g.} \cite {24}). The simplest of these equations is the
dmKP equation
\begin{equation}
V_{t}+\frac{3}{2}V^{2}V_{x}-\frac{3}{4}V_{x}\partial _{x}^{-1}V_{y}-\frac{3}{%
4}\partial _{x}^{-1}V_{yy}=0\qquad .  \label{5.13}
\end{equation}
Analogously to the KP case, one can construct also the hierarchy
of linear problems for the function $\varphi $. The simplest of
them is of the form
\begin{equation}
\frac{\partial \varphi }{\partial T_{2}}-\left( 2\frac{\partial
S}{\partial
T_{1}}+V\right) \frac{\partial \varphi }{\partial T_{1}}-\frac{\partial ^{2}S%
}{\partial T_{1}^{2}}\,\varphi =0\qquad .  \label{5.14}
\end{equation}
Analogously to the dKP case equations (\ref{5.12}) imply the
hierarchy of equations for $S$ only.

\section{Dispersionless two-dimensional Toda lattice (2DTL) hierarchy}

\setcounter{equation}{0}

Our third example is the d2DTL hierarchy. In this case $S(\lambda ;X,Y,T)$ $%
=T\ln \lambda +\sum_{n=1}^{\infty }\lambda
^{n}X_{n}+\sum_{n=1}^{\infty }\lambda ^{-n}Y_{n}$ and the domain
$G$ is the ring $D_{a,b}$ ($a\leq |\lambda |\leq b$) with the
cutted piece of the real axis. The derivatives of $S$ have now
singularities both in the origin and at the infinity:

\begin{eqnarray}
\frac{\partial S}{\partial T} &=&\ln \lambda +\frac{\partial \widetilde{S}}{%
\partial T}\quad ,  \nonumber \\
\frac{\partial S}{\partial X_{n}} &=&\lambda ^{n}+\frac{\partial \widetilde{S%
}}{\partial X_{n}}\quad ,  \label{6.1} \\
\frac{\partial S}{\partial Y_{n}} &=&\lambda ^{-n}+\frac{\partial \widetilde{%
S}}{\partial Y_{n}}\qquad ,\qquad n=1,2,3,...\quad .  \nonumber
\end{eqnarray}
Since
\begin{eqnarray}
\widetilde{S}(\lambda ;X,Y,T) &=&1+\frac{\widetilde{S}_{1}}{\lambda }+\frac{%
\widetilde{S}_{2}}{\lambda ^{2}}+...\qquad ,\quad \quad \;\lambda
\rightarrow \infty \quad ,  \nonumber \\
&&  \label{6.2} \\
\widetilde{S}(\lambda ;X,Y,T) &=&S_{0}+\lambda S_{1}+\lambda
^{2}S_{2}+...\qquad ,\quad \lambda \rightarrow 0  \nonumber
\end{eqnarray}
one has at $\lambda \rightarrow \infty $%
\begin{eqnarray}
\frac{\partial S}{\partial X_{n}} &=&\lambda ^{n}+\frac{1}{\lambda }\;\frac{%
\partial \widetilde{S}_{1}}{\partial X_{n}}+...  \nonumber \\
\frac{\partial S}{\partial Y_{n}} &=&\lambda ^{-n}+\frac{1}{\lambda }\;\frac{%
\partial \widetilde{S}_{1}}{\partial X_{n}}+...\qquad ,\qquad n=1,2,3,...
\nonumber \\
&&  \label{6.3} \\
e^{\frac{\partial S}{\partial T}} &=&\lambda +\frac{\partial \widetilde{S}%
_{1}}{\partial T}+\frac{1}{\lambda }\;\left[ \frac{\partial \widetilde{S}_{2}%
}{\partial T}+\frac{1}{2}\left( \frac{\partial \widetilde{S}_{1}}{\partial T}%
\right) ^{2}\right] \qquad ,  \nonumber \\
e^{-\frac{\partial S}{\partial T}} &=&\frac{1}{\lambda
}-\frac{1}{\lambda ^{2}}\frac{\partial \widetilde{S}_{1}}{\partial
T}  \nonumber
\end{eqnarray}
while at $\lambda \rightarrow 0$%
\begin{eqnarray}
\frac{\partial S}{\partial X_{n}} &=&\frac{\partial S_{0}}{\partial X_{n}}%
+O(\lambda )  \nonumber \\
\frac{\partial S}{\partial Y_{n}} &=&\frac{1}{\lambda
^{n}}+\frac{\partial
S_{0}}{\partial Y_{n}}+O(\lambda )\qquad ,\qquad n=1,2,3,...  \nonumber \\
&&  \label{6.4} \\
e^{\frac{\partial S}{\partial T}} &=&\lambda e^{\frac{\partial S_{0}}{%
\partial T}}+\lambda ^{2}e^{\frac{\partial S_{0}}{\partial T}}\,\frac{%
\partial S_{1}}{\partial T}+O\left( \lambda ^{3}\right) \qquad ,  \nonumber
\\
e^{-\frac{\partial S}{\partial T}} &=&\frac{1}{\lambda
}e^{-\frac{\partial
S_{0}}{\partial T}}-\frac{\partial S_{1}}{\partial T}e^{-\frac{\partial S_{0}%
}{\partial T}}+O(\lambda )\qquad \qquad .  \nonumber
\end{eqnarray}
The required function $f_{0}$ should not have singularities  at $%
\lambda =0$ and at $\lambda =\infty $ and should vanish at
$\lambda \rightarrow
\infty $. Taking the derivatives $\frac{\partial S}{\partial T}$ $,$ $\frac{%
\partial S}{\partial X_{n}}$ , $\frac{\partial S}{\partial Y_{n}}$ and using
(\ref{6.2}), (\ref{6.3}), one finds the following two equations
\begin{eqnarray}
\frac{\partial S}{\partial Y_{1}}-Ve^{-\frac{\partial S}{\partial
T}} &=&0
\label{6.5} \\
\frac{\partial S}{\partial X_{1}}-e^{\frac{\partial S}{\partial
T}}-U &=&0 \label{6.6}
\end{eqnarray}
where $V(X,Y,T)=e^{\frac{\partial S_{0}}{\partial T}}$ and $U(X,Y,T)=-\frac{%
\partial \widetilde{S}_{1}}{\partial T}$ . The system (\ref{6.5})-(\ref{6.6}%
) is the simplest system of equations for the function $S$
associated with the d2DTL hierarchy. We note that in contrast to
the papers \cite{10} we have only one function $S$.

To extract from the above system nonlinear equations for the functions $%
V(X,Y,T)$ and $U(X,Y,T)$ we perform the expansion of the l.h.s. of equation (%
\ref{6.5}) at large $\lambda $ and of the l.h.s. of equation
(\ref{6.6}) around $\lambda =0$. The terms of the order
$\frac{1}{\lambda }$ in (\ref {6.5}) give $V=1+\frac{\partial
\widetilde{S}}{\partial Y_{1}}$ while vanishing of the zero order
terms in equation (\ref{5.6}) provides us with the equation
$\frac{\partial S_{0}}{\partial X_{1}}-U=0$. As a result, we get
the system of equations
\begin{equation}
1+\frac{\partial \widetilde{S}_{1}}{\partial Y_{1}}=e^{\frac{\partial S_{0}}{%
\partial T}}\qquad ,\qquad \frac{\partial S_{0}}{\partial X_{1}}+\frac{%
\partial \widetilde{S}_{1}}{\partial T}=0\qquad .  \label{6.7}
\end{equation}
To rewrite it in a more familiar form we introduce the function $\alpha =%
\frac{\partial S_{0}}{\partial T}$ , the differentiate twice the
first equation (\ref{6.7}) with respect to $T$ and use the second
equation (\ref {6.7}). One gets
\begin{equation}
\frac{\partial \alpha }{\partial X_{1}\partial Y_{1}}+\frac{\partial ^{2}}{%
\partial T^{2}}(e^{\alpha })=0  \label{6.8}
\end{equation}
that is the standard form of the dispersionless 2DTL equation.

It is easy to show that the formal compatibility conditions for
(\ref{6.5}), (\ref{6.6}) are equivalent to the system
\begin{equation}
V_{X_{1}}-V\,U_{T}=0\qquad ,\qquad U_{Y_{1}}+V_{T}=0  \label{6.9}
\end{equation}
which, of course, again gives rise to equation (\ref{6.8}) ($\alpha =\ln V$%
). In the form (\ref{6.9}) the 2DTL equation has been derived in
\cite{8}.

Higher equations for $S$ can be obtained analogously. Taking the times $%
X_{2} $ and $Y_{2}$ , one finds the following equations
\begin{equation}
\frac{\partial S}{\partial Y_{2}}-V_{2}e^{-2\frac{\partial S}{\partial T}%
}-V_{1}e^{-\frac{\partial S}{\partial T}}=0\quad \quad ,
\label{6.10}
\end{equation}
\begin{equation}
\frac{\partial S}{\partial X_{2}}-e^{2\frac{\partial S}{\partial T}}-U_{1}e^{%
\frac{\partial S}{\partial T}}=0\qquad \quad \;\,\,  \label{6.11}
\end{equation}
where
\begin{equation}
\begin{array}{l}
V_{2}=e^{2\frac{\partial S_{0}}{\partial T}}\quad \qquad ,\qquad V_{1}=2\;%
\frac{\partial S_{1}}{\partial T}\;e^{\frac{\partial S_{0}}{\partial T}%
}\quad , \\
U_{1}=-2\;\frac{\partial \widetilde{S}_{1}}{\partial T}\quad \quad
,\qquad U_{0}=-2\;\frac{\partial \widetilde{S}_{2}}{\partial
T}\qquad \quad .
\end{array}
\label{6.12}
\end{equation}
Higher d2DTL equations have, consequently, the form
\begin{eqnarray}
\frac{\partial V_{2}}{\partial X_{2}}-2\,V_{2}\,\frac{\partial U_{0}}{%
\partial T} &=&0\qquad ,  \nonumber \\
\frac{\partial V_{1}}{\partial X_{2}}-V_{1}\frac{\partial U_{0}}{\partial T}%
-2V_{2}\,\frac{\partial U_{1}}{\partial T}-U_{1}\frac{\partial V_{2}}{%
\partial T} &=&0\quad \quad ,  \nonumber \\
&&  \label{6.13} \\
\frac{\partial U_{0}}{\partial Y_{2}}+\frac{\partial }{\partial T}%
(U_{1}V_{1})+2\,\frac{\partial V_{2}}{\partial T} &=&0\qquad ,  \nonumber \\
\frac{\partial U_{1}}{\partial Y_{2}}+2\,V_{1}\,\frac{\partial V_{1}}{%
\partial T} &=&0\qquad .  \nonumber
\end{eqnarray}
The hierarchy of equations for $S$ takes the form
\begin{eqnarray}
\frac{\partial S}{\partial Y_{n}}-\sum_{k=1}^{n}V_{nk}(X,Y,T)\;e^{-k\frac{%
\partial S}{\partial T}} &=&0\quad \qquad ,  \label{6.14} \\
\frac{\partial S}{\partial X_{n}}-\sum_{k=0}^{n}U_{nk}(X,Y,T)\;e^{k\frac{%
\partial S}{\partial T}} &=&0\quad \qquad  \label{6.15}
\end{eqnarray}
where $V_{nk}$ and $U_{nk}$ ($U_{nn}=1$) are appropriate
functions. These equations provides us with the d2DTL hierarchy
for the coefficients $V_{nk}$ and $U_{nk}$ .

The formulae (\ref{6.14}), (\ref{6.15}) shows the role of the function $e^{%
\frac{\partial S}{\partial T}}$ . In the 1+1-dimensional case this
fact was first noted in the paper \cite{5}.

The d2DTL hierarchy clearly contains the dKP and dmKP hierarchies
as sub-hierarchies. The first arises if one consider only times
$X_{n}$ putting
$T_{n}=T=0$ while the dmKP hierarchy is associated only with times $Y_{n}$ ($%
X_{n}=T=0$).

Equations (\ref{6.14}) and (\ref{6.15}) imply the hierarchy of
equations for the function $S$ only. The lowest of them is of the
form
\begin{equation}
\frac{\partial ^{2}S}{\partial X_{1}\partial Y_{1}}+e^{\frac{\partial S}{%
\partial T}}\,\frac{\partial S}{\partial Y_{1}}\;\frac{\partial ^{2}S}{%
\partial T^{2}}=0\qquad .  \label{6.16}
\end{equation}

\section{Ring of symmetries for the quasi-classical $\bar{\partial}$-problem
and universal Whitham hierarchy}

\setcounter{equation}{0}

The results of the previous section demonstrate that the
symmetries of the quasi-classical $\bar{\partial}$-problem have a
rather special property. Namely, for the dKP, dmKP and d2DTL
hierarchies different symmetries $\omega _{A}=\frac{\partial
S}{\partial T_{A}}$ are connected by certain algebraic relations
(see formulae (\ref{5.3}), (\ref{5.12}) and \ref{6.14}), (\ref
{6.15})).

This property of the symmetries of the quasi-classical $\bar{\partial}$%
-problem has, in fact, a deeper background and is of general
character. This background is provided by certain theorems about
the solutions of the Beltrami equation (see \cite{28}).

Thus, let us start with the general quasi-classical
$\bar{\partial}$-problem
\begin{equation}
\frac{\partial S}{\partial \bar{\lambda}}=W\left( \lambda ,\bar{\lambda};%
\frac{\partial S}{\partial \lambda }\right)  \label{7.1}
\end{equation}
where $W\left( \lambda ,\bar{\lambda};\xi \right) $ is a certain
function and dependence of $S$ on parameters (times) is not
specified yet.

Infinitesimal symmetries $\omega $ of the problem (\ref{7.1}) are
defined by the linear Beltrami equation
\begin{equation}
\frac{\partial \omega }{\partial \overline{\lambda }}=W^{\prime
}\left(
\lambda ,\bar{\lambda};\frac{\partial S}{\partial \lambda }\right) \;\frac{%
\partial \omega }{\partial \lambda }\qquad .  \label{7.2}
\end{equation}
Linear Beltrami equation possesses a number of interesting
properties. They have been studied in details as a part of the
theory of generalized analytic functions (see \cite{28}). The
first important property is formulated in the section 3 (Chapter
II) of the book \cite{28}. This \textbf{Theorem 1} (see the
appendix) states that for measurable and bounded on the entire
complex plane $\mathbb{C}$ functions $W^{\prime }$ which satisfies
the condition $|W^{\prime }|\leq W_{0}<1$ and some other mild
conditions, equation (\ref{7.2}) has a solution $\omega
_{0}(\lambda )$ (so-called, basic homeomorphism) for which
\begin{equation}
\omega _{0}(\lambda )=\lambda +O\left( \frac{1}{\lambda }\right)
\qquad ,\qquad \lambda \rightarrow \infty \qquad .  \label{7.3}
\end{equation}
Another Theorem  (the theorem 2.16 from \cite{28})(see the
appendix) says that all solutions (in some class) of equation
(\ref{7.2}) are given by the formula
\begin{equation}
\omega (\lambda ,\overline{\lambda })=\Omega \left( \omega _{0}(\lambda ,%
\overline{\lambda })\right)  \label{7.4}
\end{equation}
where $\Omega (\xi )$ is an arbitrary analytic function in the domain $%
\omega (D_{0})$.

These two basic results allow us to construct infinite hierarchy
associated with the problem (\ref{7.1}). Indeed, let us assign the
time $t_{A}$ for
each symmetry $\omega _{A}$ such that $\omega _{A}=\frac{\partial S}{%
\partial T_{A}}$ . So for the basic solution (\ref{7.3}) $\omega _{0}=\frac{%
\partial S}{\partial T_{0}}$ . The first theorem now says that there exists
a symmetry of equation (\ref{7.1}) such that
\begin{equation}
\frac{\partial S}{\partial T_{0}}=\lambda +O\left(
\frac{1}{\lambda }\right) \qquad .  \label{7.5}
\end{equation}
Then the second theorem states that for any symmetry $\frac{\partial S}{%
\partial T_{A}}$ one has
\begin{equation}
\frac{\partial S}{\partial T_{A}}=\Omega _{A}\left( \frac{\partial S}{%
\partial T_{0}}(\lambda ,\overline{\lambda }),T\right)  \label{7.6}
\end{equation}
where $\Omega (\xi ,T)$ is an appropriate function of the first
argument.
Thus the above two theorems imply that under certain conditions the $\bar{%
\partial}$-equation (\ref{7.1}) possesses an infinite ring of symmetries
(deformations) given by the equations
\begin{equation}
\frac{\partial S(\lambda ,\overline{\lambda };T)}{\partial
T_{A}}=\Omega _{A}\left( \frac{\partial S}{\partial
T_{0}},T\right) \qquad \quad ,\qquad A=0,1,2,3,...  \label{7.7}
\end{equation}
where $\Omega _{A}(\xi ,T)$ are arbitrary analytic functions of
the $\xi $. The set of the equations (\ref{7.7}) is compatible by
construction. Equations (\ref{7.7}) give rise to certain nonlinear
equations for functions $U_{k}(T)$ on which $S$ may depend. These
equations can be obtained also from the equations for $\Omega
_{A}$ which follow from (\ref{7.7}). They are
\begin{equation}
\frac{\partial \Omega _{A}}{\partial T_{B}}-\frac{\partial \Omega _{B}}{%
\partial T_{A}}+\left\{ \Omega _{A},\Omega _{B}\right\} =0\qquad ,\qquad
A,B=0,1,2,...  \label{7.7bis}
\end{equation}
where

\begin{equation}
\{f,g\}=\frac{\partial f}{\partial p}\,\,\frac{\partial g}{\partial T_{1}}-\,%
\frac{\partial f}{\partial T_{1}}\,\,\frac{\partial g}{\partial
p}\qquad . \label{7.8}
\end{equation}
and we denoted $p=\frac{\partial S}{\partial T_{0}}$ .

So we constructed an integrable hierarchy of equations out of the $\bar{%
\partial}$-problem (\ref{7.1}). It is an infinite ring since $\Omega
_{A}(\xi ,T)$ are arbitrary analytic functions. To get a concrete
hierarchy
one has to specify the set of functions $\Omega _{A}$. The set of functions $%
\Omega _{A}\left( p(\lambda ,\overline{\lambda };T),T\right) $ such that $%
\Omega _{k}\sim \lambda ^{k}+O\left( 1/\lambda \right) $ as
$\lambda \rightarrow \infty $ with identification $T_{0}=T_{1}$,
$T_{k}=T_{A-1}$ gives rise to the dKP hierarchy.

In the construction given above the time $T_{0}$ has played a
special role being connected with the ''basic'' symmetry $\omega
_{0}$. Infinite ring of symmetries for the problem (\ref{7.1})
admits more general and symmetric formulation. It is due to the
already mentioned obvious fact that any
differentiable function $f\left( \frac{\partial S}{\partial T_{A_{1}}},\frac{%
\partial S}{\partial T_{A_{2}}},...\right) $ of any set of symmetries $\frac{%
\partial S}{\partial T_{A_{1}}},\frac{\partial S}{\partial T_{A_{2}}}$ , ...
is again a symmetry (\textit{i.e. }a solution of equation
(\ref{7.2})). Then
the implicit function theorem implies that any symmetry $\frac{\partial S}{%
\partial T_{A}}$ can be chosen as a basic one.

So, let us take (arbitrary) symmetry $p=\frac{\partial S}{\partial
T_{0}}$ . The infinite hierarchy of symmetries now take the form
\begin{equation}
\frac{\partial S}{\partial T_{A}}=\Omega _{A}(p,T)  \label{7.9}
\end{equation}
where $\Omega _{A}(\xi ,T)$ are arbitrary differentiable functions of $\xi $%
. The compatibility conditions for equation (\ref{7.9}) is of the
form (\ref {7.7bis}), (\ref{7.8}) where now $T_{0}$ and
$p=\frac{\partial S}{\partial T_{0}}$ are arbitrary time and the
corresponding symmetry.

The infinite set of equations (\ref{7.7bis}) in this case is
nothing but the universal Whitham hierarchy introduced in the
different way in \cite{8}. So in our approach the universal
Whitham hierarchy is an infinite ring of
symmetries of the general quasi-classical $\bar{\partial}$-problem (\ref{7.1}%
).

\section{Dispersionless hierarchies of the B type}

\setcounter{equation}{0}

Various type of reductions for the dKP hierarchy have been
considered in \cite{5}, \cite{10}. Here we will discuss the
dispersionless hierarchies of the so-called B type. The
dispersionless BKP hierarchy has been discussed briefly in the
paper \cite{30}. The dBKP hierarchy is characterized by the
constraint \cite{30}
\begin{equation}
S(-\lambda ,T)=-S(\lambda ,T)\qquad .  \label{8.1}
\end{equation}
This constraint immediately implies that only odd powers of $\frac{\partial S%
}{\partial T_{1}}$ are allowed in the equations (\ref{5.3}). Since
in this case $S_{0}(\lambda ,T)=\lambda \,T_{1}+\lambda
^{3}T_{3}+\lambda
^{5}T_{5}+...$ and $\widetilde{S}=1+\frac{S_{1}}{\lambda }+\frac{S_{3}}{%
\lambda ^{3}}+...$ as $\lambda \rightarrow \infty $ , the
hierarchy of equations for $S$ takes the form
\begin{equation}
\frac{\partial S}{\partial T_{2n+1}}-\left( \frac{\partial S}{\partial T_{1}}%
\right) ^{2n+1}-\sum_{k=0}^{n-1}U_{nk}(T)\;\left( \frac{\partial
S}{\partial T_{1}}\right) ^{2k+1}=0\qquad .  \label{8.2}
\end{equation}
The two lowest equations (\ref{8.2}) are
\begin{equation}
\frac{\partial S}{\partial T_{3}}-\left( \frac{\partial S}{\partial T_{1}}%
\right) ^{3}-U\;\frac{\partial S}{\partial T_{1}}=0\qquad \qquad
\qquad , \label{8.3}
\end{equation}
\begin{equation}
\frac{\partial S}{\partial T_{5}}-\left( \frac{\partial S}{\partial T_{1}}%
\right) ^{5}-V_{3}\;\left( \frac{\partial S}{\partial
T_{1}}\right) ^{3}-V_{1}\;\frac{\partial S}{\partial T_{1}}=0\quad
\label{8.4}
\end{equation}
where
\begin{equation}
U=-3\frac{\partial S_{1}}{\partial T_{1}}\quad ,\quad V_{3}=\frac{5}{3}%
U\quad ,\quad V_{1}=\frac{5}{9}U^{2}-\frac{\partial S_{3}}{\partial T_{1}}%
\qquad .  \label{8.6}
\end{equation}
Equations (\ref{8.3}), (\ref{8.4}) implies that
\begin{equation}
\frac{9}{5}\frac{\partial U}{\partial T_{5}}+U^{2}\frac{\partial
U}{\partial
T_{1}}-U\frac{\partial U}{\partial T_{3}}-\frac{\partial U}{\partial T_{1}}%
\;\partial _{T_{1}}^{-1}\left( \frac{\partial U}{\partial
T_{3}}\right)
-\partial _{T_{1}}^{-1}\left( \frac{\partial ^{2}U}{\partial T_{3}^{2}}%
\right) =0  \label{8.7}
\end{equation}
Equation (\ref{8.6}) is the dispersionless limit of the
2+1-dimensional Sawada-Kotera (and also Kaup-Kupershmidt) equation
\cite{31},\cite{32}.

To get the d2DTL hierarchy of the B type we shall use the
universal Whitham
hierarchy equation (\ref{7.9}) and (\ref{7.7}). Due to constraint (\ref{8.1}%
) only odd functions $\Omega _{A}(-p,T)=-\Omega (p,T)$ are
admissable.
Taking the time $t_{1}=X$ as the reference one (\textit{i.e.} $p=\frac{%
\partial S}{\partial X}$) and two other equations (\ref{7.9}) in the form
\begin{equation}
\frac{\partial S}{\partial Y}=\frac{V}{p-U}-\frac{V}{p+U}\qquad
,\qquad \frac{\partial S}{\partial T}=\ln \frac{p-U}{p+U}
\label{8.8}
\end{equation}
where $u$ and $V$ are functions of $X,Y,T$ , one obtains the
equations
\begin{equation}
\begin{array}{l}
V_{T}+U_{Y}=0\qquad \qquad \;\;, \\
\qquad \\
U_{T}+\frac{U_{X}}{U}-\frac{V_{X}}{V}=0\qquad .
\end{array}
\label{8.9}
\end{equation}
Introducing the function $\beta =\ln \left( \frac{V}{U}\right) $,
one can rewrite the system (\ref{8.9}) as
\begin{equation}
\begin{array}{l}
\beta _{XY}+\left( Ue^{\beta }\right) _{TT}=0\quad \quad , \\
\\
\beta _{X}+U_{T}=0\qquad \qquad \quad .
\end{array}
\label{8.10}
\end{equation}
It is the d2DTL equation of the B type. The analog of equation
(\ref{6.5}), \ref{6.6}) for the B-d2DTL equation (\ref{8.10}) is
rather interesting
\begin{equation}
\begin{array}{l}
\frac{\partial S}{\partial Y}+\frac{V}{U}\;sh\left( \frac{\partial S}{%
\partial T}\right) =0\quad \qquad , \\
\\
\frac{\partial S}{\partial X}+U\;cth\left( \frac{1}{2}\,\frac{\partial S}{%
\partial T}\right) =0\qquad .
\end{array}
\label{8.11}
\end{equation}
The compatibility condition for this system is equivalent to the
system (\ref {8.9}).

Note finally that the nonlinear equation for $S(z,\bar{z};X,Y,T)$
in this case is of the form
\begin{equation}
\frac{S_{TT}\,S_{X}\,S_{Y}}{1+ch(S_{T})}-\frac{S_{TX}\,S_{Y}}{sh(S_{T})}%
+S_{XY}=0\qquad .  \label{8.12}
\end{equation}

\textbf{APPENDIX}

\appendix
\setcounter{equation}{0}
\renewcommand{\theequation}{A.\arabic{equation}}

There is a well established theory of generalized solutions of the
linear Beltrami equation ( see for instance \cite{21}-\cite{23}
and \cite{28})
\begin{equation}\label{a}
Z_{\bar{\lambda}}=A Z_{\lambda},
\end{equation}
where $A$ is any given measurable function $||A||_{\infty}<1$ on
$G$. Obviously, for $A\equiv 0$ we get into the class of conformal
mappings.  To present these results we need to introduce the
operators
\[
Th(\lambda):=\frac{1}{2\pi\imath}
\int\!\!\int_{\mathbb{C\,}\times\mathbb{C\,}}
\frac{h(\lambda')}{\lambda'-\lambda} \mathrm{d} \lambda' \wedge
 \bar{\lambda}',\quad
\Pi h(\lambda):=\frac{\partial Th}{\partial \lambda}(\lambda),
\]
where the integral is taken in the sense of the Cauchy principal
value. Then one has:

\begin{lemma}
For any $p>1$ the operator $\Pi$ defines a bounded operator in
$L^p(\mathbb{C\,})$ and for any $0\leq k<1$ there exists
$\delta>0$ such that
\[
k||\Pi||_p<1,
\]
for all $|p-2|<\delta$.
\end{lemma}

The next theorem summarizes the properties of solutions of
\eqref{a} that we need in our  discussion.

\begin{theorem} Given a measurable function $A$ with compact support
inside the circle $|\lambda|<R$ and such that
$||A||_{\infty}<k<1$. Then, for any fixed exponent $p=p(k)>2$ such
that $k||\Pi||_p<1$, it follows that

\begin{description}
\item[1)] There is a unique function $Z_0$ on $\mathbb{C\,}$ with distributional
derivatives satisfying the Beltrami equation \eqref{a} such that
\begin{equation}\label{z}
Z_0(\lambda)=\lambda+O(\frac{1}{\lambda}),\quad
\lambda\rightarrow\infty,
\end{equation}
with $Z_{0,{\bar \lambda}}$ and $Z_{0,\lambda}-1$ being elements
of $L^p(\mathbb{C\,})$.
\item[2)] Every solution
of \eqref{z} on a domain $G$ of  $\mathbb{C\,}$ can be represented
as
\begin{equation}\label{phi}
Z(\lambda)=\Phi(Z_0(\lambda)),
\end{equation}
where $\Phi$ is an arbitrary analytic function on the image domain
$Z_0(G)$ of $G$ under $Z_0$.
\end{description}
\end{theorem}

\textbf{Acknowledgment.} B.K. is grateful to the Departamento de
Fisica Teorica II Universidad Complutense for a kind hospitality
during which this paper was basically written.

\end{document}